# Kretschmann Invariant and Relations between Spacetime Singularities, Entropy and Information

Ioannis Gkigkitzis[1], Ioannis Haranas[2], Omiros Ragos[3]

[1]Departments of Mathematics, East Carolina University

124 Austin Building, East Fifth Street, Greenville

NC 27858-4353, USA, E-mail: gkigkitzisi@ecu.edu

[2]Department of Physics and Astronomy, York University

4700 Keele Street, Toronto, Ontario, M3J 1P3, Canada

E-mail:yiannis.haranas@gmail.com

[3]Department of Mathematics, University of Patras.

Faculty of Sciences, University of Patras, GR-26504 Patras, Greece,

Email: ragos@math.upatras.gr

**Abstract**

Using a Yukawa type of metric we derive the Kretschmann scalar (KS) for a general static black hole of mass *M*. The scalar gives the curvature of the spacetime as a function of the radial distance *r* in the vicinity as well as inside of the black hole. Furthermore, the Kretschmann scalar helps us understand the black hole's appearance as a "whole entity". It can be applied in solar mass size black holes, neutron stars or supermassive black holes at the center of various galaxies. In an effort to investigate the connection of geometry to entropy and information, the Kretschmann scalar for a solar mass Yukawa Schwarzschild and simple Schwarzschild black holes are derived. Moreover, the curvature's dependence on the entropy and number of information *N* in nats is derived.

## 1. Introduction

When we study any space time, it is important above other things to know whether the spacetime is regular or not. By regular spacetime we simply mean that the space time must have regular curvature invariants are finite at all space time points, or contain curvature singularities at which at least one such singularity is infinite. In many cases one of the most useful ways to check that is by checking for the finiteness of the Kretschmann scalar (from then on KS) which sometimes is also called Riemann tensor squared, in other words:

$$K = R_{\alpha\beta\gamma\delta}R^{\alpha\beta\gamma\delta}, \tag{1}$$

where $R_{\alpha\beta\gamma\delta}$ is the Riemann tensor. In principle the derivation of the KS is simple, but in practice to actually derive it requires a very long algebraic computation, which can very much be simplified with today's software that perform algebraic and tensorial calculations. In order to calculate the above scalar we first need to calculate the Christoffel symbols of the second kind according to the equations:

$$\Gamma^{\alpha}_{\beta\gamma} = \frac{1}{2}g^{\delta\alpha}\left(\frac{\partial g_{\gamma\delta}}{\partial x^{\beta}} + \frac{\partial g_{\beta\delta}}{\partial x^{\gamma}} + \frac{\partial g_{\gamma\beta}}{\partial x^{\delta}}\right), \tag{2}$$

once the Christoffel symbols are calculated we then calculate the Riemann tensor to be:

$$R^{\alpha}_{\beta\gamma\delta} = \frac{\partial \Gamma^{\alpha}_{\beta\delta}}{\partial x^{\gamma}} - \frac{\partial \Gamma^{\alpha}_{\beta\gamma}}{\partial x^{\delta}} + \Gamma^{\mu}_{\beta\delta}\Gamma^{\alpha}_{\mu\gamma} - \Gamma^{\mu}_{\beta\gamma}\Gamma^{\alpha}_{\mu\delta}. \tag{3}$$

For example in a sphere there are only two nonzero Riemann tensor components i.e.: $R^{1}_{212} = \sin^2\theta$ and also $R^{1}_{221} = \sin^2\theta$, which exactly characterize the curvature of the sphere. We usually think of the curvature as the Ricci scalar, which can be obtained by contraction of the Riemann tensor, first $R_{\beta\delta} = R^{\alpha}_{\beta\alpha\delta}$ and then $R = R^{\alpha}_{\alpha}$. In the case of a sphere the takes the form (Henry 2000):

$$K_s = R_{\alpha\beta\gamma\delta}R^{\alpha\beta\gamma\delta} = \frac{4}{a^4} \tag{4}$$

Because it is a sum of squares of tensor components, this is a quadratic invariant. In the case of black holes the calculation of the scalar is required if somebody wants to derive and investigate the curvature of a black hole. The need for calculation of the KS emanates from the fact that in vacuum the field equations of general relativity a zero Gaussian curvature at and in the black hole, thus giving no information about curvature of the spacetime, and thus the $K$ scalar need to be computed. In the case of Schwarzschild black holes the KS is (D'Inverno 1992):

$$K_{bh} = \frac{48G^2M^2}{c^4r^6}. \tag{5}$$

In this contribution we examine the KS of a Yukawa type modified Schwarzschild black holes as it is given in (Haranas and Gkigkitzis 2013) and compare this to the Schwarzschild scalar. Furthermore, in an effort to investigate the relation between entropy, information and geometry. For that we write the Yukawa black hole scalar as a function of entropy and information number $N$.

## 2. The Yukawa potential

Following (Capozziello, Cristofano et al. 2010) and (Haranas and Gkigkitzis 2013) we say that theories derived from the action:

$$A = \int_V \sqrt{-g}\left( f(R, \nabla R, \nabla^2 R, \nabla^k R, \phi) - \frac{\varepsilon}{2} g^{\mu\nu}\phi_{;\mu}\phi_{;\nu} + L_m \right) d^4 x, \tag{6}$$

result to Yukawa corrections to the gravitational potential where $f(R)$ is an analytic function of Ricci scalar, $g$ is the determinant of the metric $g_{\mu\nu}$, and $L_m$ is a fluid-matter Lagrangean, and where $f$ is an unspecified function of curvature invariants $R$ and $\nabla R$ and of scalar fields $\phi$. Actually, more complicated invariants like $R^{\mu\nu}R_{\mu\nu}$ and $R^{\mu\nu\alpha\beta}R_{\mu\nu\alpha\beta}$ and $C^{\mu\nu\alpha\beta}C_{\mu\nu\alpha\beta}$. In the weak field limit theories described by Eq. (6) result to potentials of the form (Capozziello, Cristofano et al. 2010):

$$V(r) = -\frac{G_\infty M}{r}\left[ 1 + \sum_{k=1}^n \alpha_k e^{-\frac{r}{\lambda_k}} \right] \tag{7}$$

where $G_\infty$ is the value of the gravitational constant as measured at infinity, $\lambda_k$ is the interaction length of the $k^{th}$ component of non-Newtonian corrections, and $\alpha_k$ amplitude of is term is normalized to the standard Newtonian term. If somebody considers the first term in the series of Eq. (9), we obtain the following potential:

$$V(r) = -\frac{G_\infty M}{r}\left( 1 + \alpha_1 e^{-\frac{r}{\lambda_1}} \right). \tag{8}$$

The second term is a Yukawa type of correction to the Newtonian potential, and its effect can be parameterized by $a_1$ and $\lambda_1$ which for simplicity we will call $\alpha$ and $\lambda$. For large distances, i.e. $r \gg \lambda$, the exponential term vanishes and the gravitational constant is simply $G_\infty$. If $r \ll \lambda$ the exponential becomes unity.

## 3. The metric and Ricci tensor and the Kretschmann scalar

Writing the metric of a spherically symmetric star in the following way:

$$ds^2 = c^2 e^{2\mu(r,t)} dt^2 - e^{2\nu(r,t)} dr^2 - e^{2\xi(r,t)} d\Omega^2 \tag{9}$$

where $\mu, \nu, \xi$ are in general functions of radial distance $r$ and time $t$ and $d\Omega^2 = d\theta^2 + \sin^2\theta d\phi^2$. For a static metric in Eq. (9) the Ricci tensor is diagonal and therefore we obtain the following nonzero components:

$$R_0^0 = e^{-2\mu}\left(2\ddot{\xi} + \ddot{\nu} + 2\dot{\xi}^2 + \dot{\nu}^2 - \dot{\mu}(2\dot{\xi} + \dot{\nu})\right) - e^{-2\nu}\left(\mu'' + \mu'(2\xi' + \mu' - \nu')\right) \quad, \tag{10}$$

$$R_1^1 = e^{-2\mu}\left(\ddot{\nu} + \dot{\nu}(2\dot{\xi} - \dot{\mu} + \dot{\nu})\right) - e^{-2\nu}\left(2\xi'' + \mu'' + 2\xi'^2 + \mu'^2 - \nu'(2\xi' + \mu')\right) \quad, \tag{11}$$

$$R_2^2 = R_3^3 = e^{-2\xi} + e^{-2\mu}\left(\ddot{\xi} + \dot{\xi}(2\dot{\xi} - \dot{\mu} + \dot{\nu})\right) - e^{-2\nu}\left(\xi'' + \xi'(2\xi' + \mu' - \nu')\right) \quad, \tag{12}$$

$$R_{01} = 2\left(\dot{\xi}' + \xi'(\dot{\xi} - \dot{\nu}) - \mu'\dot{\xi}\right). \tag{13}$$

Similarly, we find that:

$$K_1 = -R_{01}^{01} = e^{-(\mu+\nu)}\frac{d}{dr}\left(\frac{d\mu}{dr}e^{\mu-\nu}\right) \quad, \tag{14}$$

$$K_2 = -R_{02}^{02} = -R_{03}^{03} = e^{-(2\nu)}\frac{d\xi}{dr}\frac{d\mu}{dr} \quad, \tag{15}$$

$$K_3 = -R_{12}^{12} = -R_{13}^{13} = e^{-(\nu+\xi)}\frac{d}{dr}\left(e^{\xi-\nu}\frac{d\xi}{dr}\right) \quad, \tag{16}$$

$$K_4 = -R_{23}^{23} = -e^{-(2\xi)} + e^{-(2\nu)}\left(\frac{d\xi}{dr}\right)^2 \quad, \tag{17}$$

and therefore the KS becomes:

$$K = 4(K_1)^2 + 8(K_2)^2 + 8(K_3)^2 + 4(K_4)^2 \tag{18}$$

Next in our effort to investigate KS singularities of various metric let us now proceed with a fifth force metric that incorporates a Yukawa correction as it is given by (Spallicci 1991), curved only in the time coordinate, namely:

$$ds^2 = c^2\left[1 - \frac{2GM}{c^2 r}\left(1 + \alpha e^{-r/\lambda}\right)\right]dt^2 - dr^2 - r^2 d\Omega^2, \tag{18}$$

And therefore

$$\nu(r) = 0, \tag{19}$$

$$\xi(r) = r, \tag{20}$$

$$\mu(r) = \frac{1}{2}\ln\left[\left[1 - \frac{2GM}{c^2 r}\left(1 + \alpha e^{-\frac{r}{\lambda}}\right)\right]\right]. \tag{21}$$

Omitting orders $O(c^{-8})$, $O(c^{-6})$, and $e^{-3r/\lambda}$ and $e^{-4r/\lambda}$ the final expression for the KS becomes:

$$K \cong \frac{48G^2 M^2}{c^4 r^6 \lambda^2}\left[\frac{1}{2} + \alpha e^{-\frac{r}{\lambda}}\left(1 + \frac{r}{\lambda} + \frac{r^2}{3\lambda^2}\right) + \alpha^2 e^{-\frac{2r}{\lambda}}\left(\frac{1}{2} + \frac{r}{\lambda} + \frac{5r^2}{6\lambda^2} + \frac{r^3}{3\lambda^3} + \frac{r^4}{12\lambda^4}\right)\right] \tag{22}$$

Where,

$$\Lambda = \left[1 - \frac{2GM}{c^2 r}\left(1 + \alpha e^{-\frac{r}{\lambda}}\right)\right] \quad . \tag{23}$$

Following (Haranas and Gkigkitzis 2012, Haranas and Gkigkitzis 2013) we write the metric of a Yukawa type of Schwarzschild black hole that is curved in along both the time and radial coordinate, according to the equation:

$$ds^2 = c^2\left[1 - \frac{2GM}{c^2 r}\left(1 + \alpha e^{-r/\lambda}\right)\right]dt^2 - \left[1 - \frac{2GM}{c^2 r}\left(1 + \alpha e^{-r/\lambda}\right)\right]^{-1} dr^2 - r^2 d\Omega^2 \tag{24}$$

where $d\Omega^2 = d\theta^2 + \sin^2\theta\, d\phi^2$, and upon comparing Eq. (9) and (14) we obtain that:

$$v(r) = \frac{1}{2}\ln\left[\left[1 - \frac{2GM}{c^2 r}\left(1 + \alpha e^{-\frac{r}{\lambda}}\right)\right]^{-1}\right], \tag{25}$$

$$\xi(r) = r, \tag{26}$$

$$\mu(r) = \frac{1}{2}\ln\left[\left[1 - \frac{2GM}{c^2 r}\left(1 + \alpha e^{-\frac{r}{\lambda}}\right)\right]\right]. \tag{27}$$

Therefore using Eq. (18) the final expression for the KS becomes:

$$K_{Yuk} = K_{sch} + K_{cor} = \frac{48 G^2 M^2}{c^4 r^6}\left[1 + \alpha e^{-\frac{r}{\lambda}}\left(1 + \frac{16r}{12\lambda} + \frac{r^2}{\lambda^2}\right) + \alpha^2 e^{-\frac{2r}{\lambda}}\left(1 + \frac{16r}{12\lambda} + \frac{r^2}{\lambda^2} + \frac{r^3}{3\lambda^3} + \frac{r^4}{12\lambda^4}\right)\right]. \tag{28}$$

As a check we see that when $\alpha = 0$ we immediately obtain the KS of a Schwarzschild metric as it is given in (Henry 2000).

Next, we use the non-rotating but electrically charged source vacuum solution as it is given by the Reissner – Nordstrom solution (Misner, Thorne et al. 1973):

$$ds^2 = c^2\left[1 - \frac{2GM}{c^2 r} + \frac{GQ^2}{4\pi\varepsilon_0 c^4 r^2}\right]dt^2 - \left[1 - \frac{2GM}{c^2 r} + \frac{GQ^2}{4\pi\varepsilon_0 c^4 r^2}\right]^{-1} dr^2 - r^2 d\Omega^2 \; . \tag{29}$$

The final expression for the KS becomes:

$$K = \frac{48 G^2 M^2}{c^4 r^6}\left(1 + \frac{2Q^2}{c^2 rM} + \frac{7Q^4}{48 c^4 r^2 M^2}\right). \tag{30}$$

Next, let us consider the internal star metric solution that reads ():

$$ds^2 = c^2\left[\frac{3}{2}\sqrt{1 - \frac{2GM}{Rc^2}} - \frac{1}{2}\sqrt{1 - \frac{2GMr^2}{Rc^2}}\right]^2 dt^2 - \left[1 - \frac{2GMr^2}{c^2 R^3}\right]^{-1} dr^2 - r^2 d\Omega^2 \tag{31}$$

The final expression for the KS becomes:

$$K = \frac{48G^2M^2}{c^4R^6} \left[ \begin{array}{l} 1 + \dfrac{4G^2M^2r^4}{3\left(-2GMc^2r^2 + 3c^4R^6\sqrt{\dfrac{GMr^2}{c^2R^3}\left(2 - \dfrac{4GM}{c^2R}\right)}\right)^2} + \dfrac{4G^2M^2r^2}{3\left(c^2\sqrt{2}GMr^2R^3 - 3c^4R^6\sqrt{\dfrac{GMr^2}{c^2R^3}\left(1 - \dfrac{2GM}{c^2R}\right)}\right)^2} \\ - \dfrac{4c^2GMr^2R^3}{3\left(c^2GMR^3\sqrt{2}r^2 - 3c^4R^6\sqrt{\dfrac{GMr^2}{c^2R^3}\left(1 - \dfrac{2GM}{c^2R}\right)}\right)^2} + \dfrac{c^4GMR^6}{3\left(\sqrt{2}c^2GMr^2R^3 - 3c^4R^6\sqrt{\dfrac{GMr^2}{c^2R^3}\left(1 - \dfrac{2GM}{c^2R}\right)}\right)^2} \end{array} \right] \quad (32)$$

## 4. Kretschmann scalar and it relation to entropy and information

In general relativity there exist a set of curvature invariants that they are scalars. They can be formed from the Riemann, Ricci, and Weyl tensors respectively, and they describe various possible operations such that, covariant differentiation, contraction and dualisation. We can obtain various invariants that they are formed from these curvature tensors play an important role in the classification of spacetimes. Invariants useful in distinguishing Riemannian manifolds, or manifolds that they have a positive and well defined metric tensor. In order to investigate true singularities one must look at quantities that are independent of the choice of coordinates. One of these is the KS which, at the origin only or $r = 0$ possesses a singularity that is a physical singularity which we can also call gravitational singularity. At $r = 0$ the curvature becomes infinite indicating the presence of a singularity. At this point the metric, and space-time itself, is no longer well-defined. For a long time it was thought that such a solution was non-physical. However, a greater understanding of general relativity led to the realization that such singularities were a generic feature of the theory and not just an exotic special case. Such solutions are now believed to exist and are termed black holes

Our paper is an effort to connect geometry to information we first start by calculating the so called KS. Since $\alpha = 3.04 \times 10^{-8}$ and $r < \lambda = 4.94 \times 10^{15}$ m (Haranas and Gkigkitzis, 2011) and considering a 1 solar mass black hole Eq. (28) the exponential quantities $e^{-r/\lambda} \cong 1$ and $e^{-2r/\lambda} \cong 1$ are both approximately equal Eq. (22) simplifies to:

$$K \cong \frac{48G^2M^2}{c^4r^6\left[1 - \dfrac{2GM}{c^2r}(1+\alpha)\right]^2} \left[\frac{1}{2} + \alpha\left(1 + \frac{r}{\lambda}\right) + \alpha^2\left(\frac{1}{2} + \frac{r}{\lambda}\right)\right], \quad (35)$$

and therefore we see that the KS has a singularity that is given by the equation:

$$r_{KS_H} \cong \frac{2GM}{c^2r}(1+\alpha). \quad (36)$$

At this point we say say this particular metric does posses a event horizon, and therefore the entropy cannot be calculated on the horizon. To proceed, let us now define a KS entropy that can be calculated using Haranas and Gkigkitzis (2011), and therefore we write:

$$S_{KS} = \left(\frac{c^3 k_B}{4G\hbar}\right) A_{KS} = \frac{4\pi(1+\alpha)^2 k_B G}{c\hbar} M^2 = 4\pi(1+\alpha)^2 k_B \left(\frac{M}{m_p}\right)^2, \qquad (37)$$

From which we obtain that:

$$M = \frac{m_p}{2(1+\alpha)} \sqrt{\frac{S_{KS}}{\pi k_B}}, \qquad (38)$$

And

$$r_{KS_H} = \frac{Gm_p}{c^2(1+\alpha)} \sqrt{\frac{S_{KS}}{\pi k_B}}. \qquad (39)$$

Substituting Eq. (38) in Eq. (39) in Eq. (35) we express KS as a function of the entropy $S_{KS}$ and we find that:

$$K_{KS}(S_{KS}) = \frac{12 c^8 k_B}{\alpha^2 G^4 m_P^4 S_{KS}^2} \left[\frac{1}{2} + \alpha\left(1 + \frac{Gm_p}{c^2 \lambda(1+\alpha)} \sqrt{\frac{S_{KS}}{k_B}}\right) + \alpha^2\left(1 + \frac{Gm_p}{c^2 \lambda(1+\alpha)} \sqrt{\frac{S_{KS}}{k_B}}\right)\right], \qquad (40)$$

For 1 solar mass black hole the most significant contribution comes from the coefficient of the $1/S^2$ term and therefore the KS entropy can be written as:

$$S_{KS} = \pm \frac{c^4 k_B (1+\alpha)^3}{\alpha G^2 m_p^2} \sqrt{\frac{6}{K_{KS}}}, \qquad (41)$$

and therefore the number of information $N$ can be written as a function of the KS in the following way.

$$N(K_{KS}) = \frac{c^4 (1+\alpha)^3}{\alpha G^2 m_p^2 \ln 2} \sqrt{\frac{6}{K_{KS}}}. \qquad (42)$$

Similarly, taking into account the same assumptions as in Eq. (35) we obtain KS of a fully curved Yukawa type of metric is given by Eq. (28) can be approximated in the following way:

$$K_{Yuk} = K_{sch} + K_{cor} \cong \frac{48 G^2 M^2}{c^4 r^6} \left[1 + \alpha\left(1 + \frac{16r}{12\lambda}\right) + \alpha^2\left(1 + \frac{16r}{12\lambda}\right)\right], \qquad (41)$$

Using expressions for the mass and the gravitational radius for a Yukawa black hole as a functions of entropy for a solar mass black hole, as it given in (Haranas and Gkigkitzis 2013), and searching for a new kind of entropic or information theoretic singularity we have:

$$M(S) = \frac{1}{2} m_P \sqrt{\frac{S}{\pi k_B (1-\alpha)}}, \qquad (33)$$

$$r_H \cong \frac{2GM}{c^2}(1+\alpha) = \frac{2G}{c^2}(1+\alpha) \frac{m_P}{2} \sqrt{\frac{(1+\alpha)S}{\pi k_B}} = \ell_P (1+\alpha) \sqrt{\frac{S}{\pi k_B (1-\alpha)}}. \qquad (34)$$

equation (41) has only one singularity at $r = 0$ and there is no other essential singularity. Using Eq. (41) and taking into account the same assumptions as in Eq. (35) we obtain that the KS as a function of entropy $S$ on the horizon of such a black hole up to the $O(\alpha^3)$ the most significant contribution comes from the coefficient of the $1/S^2$ term and therefore the KS can be written as:

$$K_{Yuk} \cong \frac{12\pi^2 G^2 m_P^2 k_B^2}{c^4 \ell_p^6 (1+\alpha)^6 S^2}\left(1-\alpha-\alpha^3\right), \tag{42}$$

which tends to zero as the entropy approaches infinity. Solving for the entropy $S$ we obtain an expression of the entropy as a function of the scalar $K_{yuk}$ to be:

$$S = \pm \frac{2\pi G m_p k_B}{c^2 \ell_P^3} \frac{(1-\alpha)}{(1+\alpha)^3}\left[\frac{3(1+\alpha+\alpha^2)}{K_{yuk}}\right]^{1/2} \tag{43}$$

Next using Haranas and Gkigkitzis (2013a) we write the entropy to be: $S = Nk_B \ln 2$, we write the KS as a function of entropy and number of information $N$ according to the equations:

$$N = \frac{2\pi G m_p}{\ln 2 c^2 \ell_p^3} \frac{(1-\alpha)}{(1+\alpha)^3}\left[\frac{3(1+\alpha+\alpha^2)}{K_{Yuk}}\right]^{1/2}. \tag{44}$$

Both equations (43) and (44) can exhibit a singularity if and only if the coupling constant becomes negative i.e $\alpha = -1$. A negative $\alpha$ would introduce a repulsive component in the Yukawa correction something that it is known as a fifth force.

Next we consider the Reissner – Nordstrom (RN) metric the KS does not possess a singularity besides the $r = 0$ case. The event horizon RN metric has the following solutions

$$r_{RN_H} = \frac{GM}{c^2} \pm \frac{1}{2\varepsilon_0 c^2}\sqrt{\frac{G\varepsilon_0\left(Q^2 + 4\pi\varepsilon_0 GM^2\right)}{\pi}}, \tag{45}$$

And therefore the corresponding entropy becomes:

$$S_{KS} = \left(\frac{c^3 k_B}{4G\hbar}\right)A_{KS} = \frac{2\pi k_B GM^2}{c\hbar} + \frac{k_B Q^2}{4\varepsilon_0 c\hbar} \pm \frac{k_B M \sqrt{\pi G\varepsilon_0 \left(Q^2 + 4\pi\varepsilon_0 GM^2\right)}}{\varepsilon_0 c\hbar}, \tag{46}$$

Solving for the mass $M$ in terms of the entropy we obtain:

$$M = \frac{1}{8}\sqrt{\frac{16 c\hbar S}{\pi G k_B} + \frac{k_B Q^4}{\pi \varepsilon_0^2 c\hbar G S} - \frac{8Q^2}{\pi \varepsilon_0 G}}, \tag{47}$$

similarly the event horizon as a function of entropy becomes

$$r_{NR_H} = \frac{1}{8c^2}\sqrt{\frac{16 G c\hbar S}{\pi k_B} + \frac{G k_B Q^4}{\pi \varepsilon_0^2 c\hbar S} - \frac{8GQ^2}{\pi \varepsilon_0}} \pm \frac{1}{2\varepsilon_0 c^2}\sqrt{\frac{\varepsilon_0}{\pi}\left(GQ^2 + \frac{c\hbar G S \varepsilon_0}{k_B} + \frac{k_B G Q^4}{16 c\hbar \varepsilon_0 S} - \frac{GQ^2}{2\varepsilon_0}\right)}, \tag{48}$$

And therefore substituting for $r_{NR_H}$ and $M$ we write the KS of the Reissner–Nordstrom (RN) metric in the following way:

$$K_{KS} = \frac{7G^2Q^4}{2\pi^2 c^8 \varepsilon_0^2 \Phi^8} + \frac{3Q^2}{c^6 \varepsilon_0 \Phi^7} \sqrt{\frac{16c\hbar G^3 S}{\pi^3 k_B} + \frac{k_B^2 G^3 Q^4}{c\hbar \varepsilon_0^2 k_B \pi^3 S} - \frac{8G^3 Q^2}{\varepsilon_0 \pi^3}} + \frac{3}{4\pi k_B c^4 \Phi^6}\left(16Gc\hbar S + \frac{k_B^2 GQ^4}{c\hbar \varepsilon_0^2 S} - \frac{8k_B GQ^2}{\varepsilon_0}\right) \quad (49)$$

Where

$$\Phi = \frac{1}{8c^2}\sqrt{\frac{16c\hbar GS^2 + \frac{k_B^2 Q^4}{c\hbar \varepsilon_0^2} - \frac{8k_B Q^2 S}{\varepsilon_0}}{\pi k_B S}} + \frac{1}{2\varepsilon_0 c^2}\sqrt{\frac{\varepsilon_0 G}{\pi}\left(Q^2 + \frac{\varepsilon_0}{16k_B S}\left(16c\hbar S^2 + \frac{k_B^2 Q^2}{c\hbar \varepsilon_0^2} - \frac{8k_B Q^2 S}{\varepsilon_0}\right)\right)}. \quad (50)$$

The KS of the Reissner–Nordstrom (RN) metric possesses a singularity for the following values of entropies if $\Phi = 0$, and therefore we find that:

$$S_1 = \frac{k_B Q^2 (2 + c\sqrt{\varepsilon_0})}{4c\hbar \varepsilon_0 (2 - c\varepsilon_0)} \quad (51)$$

$$S_2 = \frac{k_B Q^2 (2 - c\sqrt{\varepsilon_0})}{4c\hbar \varepsilon_0 (2 + c\varepsilon_0)} \quad (52)$$

$$S_3 = \frac{k_B Q^2 (2i + c\sqrt{\varepsilon_0})}{4c\hbar \varepsilon_0 (2i - c\varepsilon_0)} \quad (53)$$

$$S_4 = \frac{k_B Q^2 (2i - c\sqrt{\varepsilon_0})}{4c\hbar \varepsilon_0 (2i + c\varepsilon_0)} \quad (54)$$

Eq. (50) can be then written as a function of the information number $N$ in the following way:

$$K_{KS} = \frac{7G^2 Q^4}{2\pi^2 c^8 \varepsilon_0^2 H^8} + \frac{3Q^2 G^{3/2}}{c^6 \varepsilon_0 H^7 k_B \pi^{3/2} \sqrt{N\ln 2}} \sqrt{\frac{k_B^2 Q^4}{c\hbar \varepsilon_0^2} - \frac{8k_B^2 Q^2 N \ln 2}{\varepsilon_0} + 16c\hbar k_B^2 N^2 (\ln 2)^2}$$

$$+ \frac{3G}{4\pi c^4 k_B^2 H^6 N \ln 2}\left(\frac{k_B^2 Q^4}{c\hbar \varepsilon_0^2} - \frac{8k_B^2 Q^2 N \ln 2}{\varepsilon_0} + 16c\hbar k_B^2 N^2 (\ln 2)^2\right) \quad (55)$$

where

$$H = \frac{1}{8c^2}\sqrt{\frac{G}{N\pi \ln 2}\left[\frac{Q^4}{c\hbar \varepsilon_0^2} - \frac{8Q^2 N \ln 2}{\varepsilon_0} + 16c\hbar N^2 (\ln 2)^2\right]}^{1/2}$$

$$+ \frac{1}{2\varepsilon_0 c^2}\sqrt{\frac{G\varepsilon_0}{\pi \varepsilon_0}\left[Q^2 + \frac{\varepsilon_0}{16N\ln 2}\left(\frac{Q^4}{c\hbar \varepsilon_0^2} - \frac{8NQ^2 \ln 2}{\varepsilon_0} + 16c\hbar N^2 (\ln 2)^2\right)\right]} \quad (56)$$

Similarly KS of the Reissner–Nordstrom (RN) metric possesses a singularity for the following values of the information number if $H = 0$, and therefore we obtain:

$$N_1 = -\frac{2c\hbar \varepsilon_0 (k_B^2 Q^2 - 2c\hbar \varepsilon_0) + \left(-W + 12\sqrt{3X}\right)^{1/3} + \frac{c^2 \hbar^2 \varepsilon_0^2 L}{\left(-W + 12\sqrt{3X}\right)^{1/3}}}{12\varepsilon_0^2 c^2 \hbar^2 k_0^2 \ln 2}, \quad (57)$$

$$N_2 = \frac{4c\hbar\varepsilon_0\left(2c\hbar\varepsilon_0 - k_B^2 Q^2\right) + \left(1 - i\sqrt{3}\right)\left(-W + 12\sqrt{3X}\right)^{1/3} + \frac{\left(1 + i\sqrt{3}\right)c^2\hbar^2\varepsilon_0^2 L}{\left(-W + 12\sqrt{3X}\right)^{1/3}}}{12\varepsilon_0^2 c^2 \hbar^2 k_0^2 \ln 2}, \tag{58}$$

$$N_2 = \frac{4c\hbar\varepsilon_0\left(2c\hbar\varepsilon_0 - k_B^2 Q^2\right) + \left(1 + i\sqrt{3}\right)\left(-W + 12\sqrt{3X}\right)^{1/3} + \frac{\left(1 - i\sqrt{3}\right)c^2\hbar^2\varepsilon_0^2 L}{\left(-W + 12\sqrt{3X}\right)^{1/3}}}{12\varepsilon_0^2 c^2 \hbar^2 k_0^2 \ln 2}, \tag{59}$$

Where

$$X = \left(c^7 \hbar^7 \varepsilon^7 k_B^6 Q^6 \left(k_B^4 Q^4 + 44 c\hbar\varepsilon_0 k_B^2 Q^2 - 16 c^2 \hbar^2 \varepsilon^2\right)\right) \tag{60}$$

$$W = \left(c^3 \hbar^3 \varepsilon^3 \left(k_B^6 Q^6 + 156 c\hbar\varepsilon_0 k_B^4 Q^4 - 240 c^2 \hbar^2 \varepsilon^2 k_B^2 Q^2 + 64 c^3 \hbar^3 \varepsilon^3\right)\right) \tag{61}$$

$$L = \left(k_B^4 Q^4 - 40 c\hbar\varepsilon_0 k_B^2 Q^2 + 16 c^2 \hbar^2 \varepsilon^2\right) \tag{62}$$

To compare we write the KS of the a Schwarzschild metric on the horizon an a function of entropy and information to be:

$$K_{Sch} = \frac{12 G\hbar}{\pi c^3 k_B r^6} S = \frac{12}{\pi k_B}\left(\frac{\ell_P}{r^3}\right)^2 S, \tag{56}$$

which at the horizon $r_H = 2GM/c^2$ becomes:

$$K_{Sch} = \frac{3\ell_P^2 c^{12}}{16\pi k_B G^6 M^6} S = \frac{12 \ell_P^2}{\pi k_B R_{gr}^6} S. \tag{57}$$

And therefore Eq. (37) in terms of the number of information $N$ the KS of the a Schwarzschild metric takes the form:

$$K_{Sch} = \frac{12 \ell_P^2}{\pi R_{gr}^2} N \ln 2, \tag{58}$$

And therefore we can express the number of information $N$ at the horizon of a Schwarzschild black hole as a function of the KS of the same black hole to be:

$$N = \left(\frac{\pi R_{gr}^6}{12 \ln 2 \ell_P^2}\right) K_{Sch}. \tag{39}$$

## 5. Discussion

**Complex Entropy**

The notion of complex-valued information entropy needs to be given some physical sense interpretation through possibly an interaction range consideration. It is known that complex entropy has been associated with discrete scale. Discrete scale invariance and its associated complex exponents may appear "spontaneously" in Euclidean systems, i.e. without the need for a pre-existing hierarchy, i.e. a hierarchical system that contains by construction a discrete hierarchy of scales occurring according to a geometrical series, where one expects and does find complex exponents and their associated log-periodic structures(Sornette 1998). Examples are (Sornette 1998): fractal dimensions of Cantor sets, ultrametric structures, wave propagation in fractal systems, magnetic and resistive effects on a system of wires connected along the Sierpinski gasket, Ising and Potts models, fiber bundle ruptures, etc. It has been shown that classical close to-critical black holes (obeying Einstein's equations) coupled to a massless complex scalar field have a leading real exponent and a subleading complex exponent (Choptuik 1993),which would correspond to a log-periodic spectrum of masses. Alternatively, the real and complex exponents control the time development of the black hole instability which is also log-periodic in time, corresponding to continuous phase oscillations of the field(Sornette 1998). At the same time, typical statistical mechanics approach starts from the partition function Z defined through the sum of all the Boltzmann factors measuring the probability of occurrence of the various states of a Hamiltonian H. The partition function takes the form

$$Z_s(T) = \sum_{\nu=1}^{N_s} e^{-\beta \cdot H} \qquad (40)$$

when observing a system at a scale s and at some temperature

$$T = \frac{1}{k_B \beta}. \qquad (41)$$

The corresponding free energy is given by

$$F_s(T) = -k_B T \ln(Z_s(T)) \qquad (42)$$

This free energy is known to be a homogeneous function at critical points, in the scaleless situation(Stanley 1971) which leads to define (real) critical exponents. An attempt to a dynamical interpretation of a classical complex free energy is found in (Zwerger 1985) where the author pointed out that the problem is to determine a characteristic "relaxation" time for some process in which the dynamical (Langevin or Fokker-Planck) equation is connected to some Hamiltonian or some corresponding transfer matrix. A probability current can be written, in fact, in terms of some unstable mode times an equilibrium factor which is the imaginary part of the free energy. The imaginary part of the free energy (or largest eigenvalues) give some information about the "nucleation stage" of the dynamics, i.e. an initial decay rate out of a metastable state of a system after a sudden change in an effective field(Zwerger 1985). In fact, it has been shown that an instantaneous quench to a metastable situation is proportional to the corresponding Im F(Zwerger 1985). This metastable state here may correspond to a physical gravitational singularity, independent of coordinates, where instead of some "relaxation time", one may consider the "spatial aspect of the phenomenon", e.g. through some correlation range length ξ that corresponds to a certain differential distance of the horizon. The real part, of course of such a free energy, determines the equilibrium energy state. In fact, any (thermodynamic) property derived from

$$F_s(T) \sim F(\xi(\varepsilon)) \tag{43}$$

where

$$\varepsilon = \frac{T - T_c}{T_c} \tag{44}$$

and

$$T, T_c \tag{45}$$

are the temperature of the system and a critical value of the temperature, and $\xi$ is the coherence length which is temperature dependent and related to the scale of interest, can be complex(Rotundo and Ausloos 2013). Therefore the complex entropy may be the result of a certain background free energy but also its corrections due to some underlying scale structure. From a quantum mechanics point of view, this might indicate the presence of a non Hermitian Hamiltonian (Rotundo and Ausloos 2013).

**Negative entropy**

Finding a setting in which the entropy's full range of positive and negative values would have a meaningful interpretation is an essential step in interpreting our results of negative entropy. Negative entropy appears as an interesting feature of higher derivative gravity(Cvetic, Nojiri et al. 2002) (Schwarzschild-Anti-deSitter and Schwarzschild-deSitter gravity black holes), and has been suggested that this appearance of negative entropy may indicate a new type instability where a transition between SdS (SAdS) black holes with negative entropy to SAdS (SdS) black holes with positive entropy would occur and where the classical thermodynamics would not apply any more. It was also suggested that such a new type phase transition is presumably related with the known fact that strong energy condition in higher derivative gravity maybe violated (Cvetic, Nojiri et al. 2002). On the other hand, entropy can be negative and erasing bit from a system can result in a net gain of work, leading to the result of the effect that a quantum computer may cool itself by erasing bits of information (del Rio, Aberg et al. 2011). It will be interesting if such negative entropy idea can apply to black holes. This can change Hawking radiation – maybe under some condition, the black hole can also be cooling system (not only radiatiing) or some other interesting effect may be happen. Modeling our universe as just a big computer, deleting information in a black hole may occur through accretion matter to black hole and the black hole does not radiate but cools the universe. This seems to be in accordance with another approach of suggesting that Hawking radiation, via quantum tunneling, carries entropy out of the black hole, a process that represents a net gain of information drawn out from the black hole(Zhang, Cai et al. 2011). So Hawking radiation carries not only entropy, but also information, out of the black hole.

The negative sign of the entropy points towards a quantum entropy which can be understood operationally as quantum channel capacity(Horodecki, Oppenheim et al. 2005). In the language of communication theory, the amount of information originating from a source is the memory required to faithfully represent its output and this amount is given by its entropy. Negative entropy has a physical interpretation in terms of how much quantum communication is needed to gain complete quantum information possession of a system in the total state(Horodecki, Oppenheim et al. 2005). With quantum particles, however, it is possible for a black hole to correlate with its environment in ways such as quantum entanglement using procedures such as quantum teleportation where qubits (the basic units of quantum information) can be transmitted exactly (in principle) from one location to another, without the qubits being transmitted through the intervening space (reference from Nature). Negativity of entropy and information has also been correlated to a reference system (A classical analogue of negative information, Oppenheim) that regulates maximally entangled states of a system which can be later be used as a teleportation protocol to transmit quantum states from a source to a receiver without the use of a quantum channel.

'Erasure' of a system is defined as taking it to a pre-defined pure state, $|0\rangle$ (del Rio, Aberg et al. 2011). Landauer's principle states that the erasure of data stored in a system has an inherent work cost and therefore dissipates heat. However, this consideration assumes that the information about the system to be erased is classical, and does not extend to the general case where an observer may have quantum information about the system to be erased, for instance by means of a quantum memory entangled with the system. The work cost of erasure is determined by the entropy of the system, conditioned on the quantum information an observer has about it. In other words, the more an observer knows about the system, the less it costs to erase it. Entropies can become negative in the quantum case, when an observer who is strongly correlated with a system may gain work while erasing it, thereby cooling the environment(del Rio, Aberg et al. 2011).

## Conclusion

Black hole is perhaps the least exemplary of physical entities. Indeed, despite their appearance at the center of most if not all spiral galaxies, black holes are entities for which it is most often said "physics breaks down". These new findings may open the door to solving many previously intractable problems in quantum information theory of black holes. It is concluded that such generalizations are not only interesting and necessary for discussing information theoretic properties of black hole gravity, but also may give new insight into conceptual ideas about entropy and information.